\documentclass[prb,showpacs,twocolumn,floatfix,amsmath,amssymb,superscriptaddress,]{revtex4}

\usepackage{amsmath}

\usepackage{amssymb}

\usepackage{graphicx}

\usepackage{dcolumn}

\usepackage{bm}

\usepackage{color}

\def\be{\begin{eqnarray}}

\def\ee{\end{eqnarray}}

\def\ba{\begin{array}}

\def\ea{\end{array}}

\begin{document}


\title{An analytical model of fractional overshooting}

\author{A. Salman}
\affiliation{Physics Department, Akdeniz University, Antalya, Turkey}
\author{A. I. Mese}
\affiliation{Physics Department, Trakya University, Edirne, Turkey}
\author{M. B. Y\"ucel}
\affiliation{Physics Department, Akdeniz University, Antalya, Turkey}
\author{A. S\i dd\i ki} %
\affiliation{Physics Department, Faculty of Sciences, Istanbul University, 34134 Vezneciler-Istanbul, Turkey } \affiliation{Physics Department, Harvard University, Cambridge 02138 MA, USA}

\begin{abstract}
We predict resistance anomalies to be observed at high mobility two dimensional electron systems (2DESs) in the fractional quantized Hall regime, where the narrow ($L<10~\mu$m) Hall bar is defined by top gates. An analytic calculation scheme is used to describe the formation of integral and fractional incompressible strips. We incorporate the screening properties of the 2DES, together with the effects of perpendicular magnetic field, to calculate the effective widths of the current carrying channels. The many-body effects are included to our calculation scheme through the energy gap obtained from the well accepted formulation of the composite fermions. We show that, the fractional incompressible strips at the edges, assuming different filling factors, become evanescent and co-exist at certain magnetic field intervals yielding an overshoot at the Hall resistance. Similar to that of the integral quantized Hall effect. We also provide a mechanism to explain the absence of 1/3 state at the Fabry-Perot interference experiments. Yet, an un-investigated sample design is proposed to observe and enhance the fragile effects like interference and overshooting based on our analytical model.
\end{abstract}




\maketitle

\section{Introduction}
The interest in investigating low-dimensional charge systems is boosted by the discoveries of quantized behavior at the globally measured resistances, when these systems are subject to high perpendicular magnetic fields $B$. In particular, integral~\cite{vKlitzing80:494} and fractional~\cite{FQHE} quantized Hall effects still stay at the center of many research activities in different context.~\cite{Goldman05:155313,Sailer:10} The integral quantized Hall effect (IQHE), observed at two dimensional electron systems (2DESs), is attributed to single particle gap formed due to high magnetic field. Whereas, the fractional quantized Hall effect (FQHE) is usually discussed within the frame work of many-body interactions, i.e. the exchange and correlation effects.~\cite{Dassarma} Over the years, the role of direct Coulomb interaction (i.e. the Hartree potential) is also shown to be important for the IQHE, due to formation of compressible/incompressible strips.~\cite{Wulf88:4218,Chang90:871,Chklovskii92:4026} Here, the electronic system is composed of co-existing metal-like (compressible) and insulator-like (incompressible) states, where the Fermi energy equals to the Landau level at the former, and falls in between consecutive levels in the latter. The semi-classical transport calculations~\cite{Guven03:115327,siddiki2004} and local probe experiments~\cite{Ahlswede02:165,Yacoby04:328,Franck:10} prove that, the excess current flows from the incompressible strips by the virtue of their scattering free properties, if their widths are large enough to accommodate at least a single electron.~\cite{siddiki2004} This direct Coulomb interaction based model of the IQHE is known as the screening theory and is successfully implemented to a number of experimental systems.~\cite{Ahlswede02:165,Mares:09,friedland,jose:epl,Sailer:10,afif:expas} Most of the fractional states, can be explained in terms of the IQHE if one replaces electrons by a new particle which is composed of an even number of flux quanta attached to a single electron.~\cite{Jain89} This mapping is known as the composite fermion picture. In his pioneering work, Beenakker~\cite{Beenakker:90(FQHE)} proposed that, the fractional edge states also exist however are very different in their nature compared to the IQHE edge-states. In a later work, Chklovskii et al.~\cite{Chklovskii92:4026} extended their electrostatic picture to show that incompressible strips emerge also in the fractional domain. There, in contrast to IQHE, the properties of the strips are determined by the many-body effects. The formation and dynamics of the fractional edge states are investigated in many experiments, of which we mention only two, the magneto-capacitance~\cite{oto:97} and the edge magneto-plasmon~\cite{Ernst:97} measurements, that are related with our discussion. In magneto-capacitance experiments it is shown that, the widths of the fractional incompressible strips are wider than expected single particle picture of the IQHE. In the edge magneto-plasmon experiments, characteristic deviations at the 2/3 state is reported, where a large sample is used with top gate. The estimates of the incompressible strip widths and the density distribution profile is also provided. These experiments both correspond to a smooth edge profile. In addition, theoretical semiclassical calculations including many-body effects within a Thomas-Fermi approximation provide plausible estimates both for the widths of the incompressible strips and edge magneto-plasmon velocities. However, this approach is mainly limited to smooth edges.

In this communication, we employ the non-self consistent picture of Chklovskii et al., together with corrections to electrostatics and wavefunctions, to investigate the co-existence of more than one fractional state at a given magnetic field interval. Such a co-existence of decaying (evanescent) incompressible strips is already shown to result in anomalous increase of transverse resistance in the integral domain.~\cite{Sailer:10,Overshoot:theo} First, we re-introduce the electrostatic calculation scheme briefly. Then a modification to the charge density profile is introduced, which is motivated by self-consistent calculations.~\cite{Sefa08:prb} Next, we use the results of composite fermion picture to determine the energy gaps induced by many-body interactions considering certain fractional states. We utilize these gaps to investigate the existence of incompressible strips of the corresponding states. Focusing on the edge properties of the system, we show that the fractional strips collapse at the samples which exhibit steep density variation. The collapse is observed once their widths become narrower than the cyclotron radius of the composite fermion associated. Despite this fact, if the edge density profile is sufficiently smooth such that more than one evanescent incompressible state can co-exist, we predict that an increase in the transverse resistance $R_{xy}$ should be observed. First, we resemble this effect with the overshooting of the integer quantized Hall effect~\cite{Sailer:10,Overshoot:theo} and claim that by the help of additional side gates one may enhance the visibility of the measurements by a considerable amount. Second, we show that the recent experimental reports on the local measurements of the energy gaps can be understood within our approach.~\cite{Deviatov09:spe,Deviatov09:fabry} There, it is found that the principle Laughlin incompressible strip is missing at the interference pattern. We also relate our findings with the particle and the quasi-particle interference experiments, where the Aharonov-Bohm phase of the fractional states are investigated.~\cite{camino05:075342, camino06:115301}
\section{The electrostatic model with interactions}
In this section we represent the essential findings of the non-self-consistent calculation scheme.~\cite{Chklovskii92:4026} Due to electrostatic stability arguments, it was proposed that the direct Coulomb interaction should modify the implausible step-like density profile resulting from non-interacting Landauer-Buettiker picture.~\cite{Buettiker86:1761} In addition, it was shown that the electron density is kept constant within a narrow strip in the close vicinity of the integral (and the fractional) filling factors $\nu=2\pi \ell_B^2 n_{\rm el}$, where $\ell_B=\sqrt{\hbar/eB}$ is the magnetic length and $n_{\rm el}$ is the electron density. These regions are called incompressible, since all the levels below the Fermi energy are fully occupied and a discontinuity is encountered if one calculates the variation of the particle number $N$ with respect to the chemical potential $\mu$. The widths of these regions are determined by the strength of the interactions together with the electrostatic boundary conditions.~\cite{Lier94:7757} Different boundary conditions are discussed in the literature, however, the Chklovskii and the Gefland~\cite{Halperin94:etchedge} models are the most known. The former considers a gate defined Hall bar, whereas the latter takes into account the chemical etching process. Both models assume translational invariance in the current direction $y$ and the 2DES to be confined in $x$ due to remote donors. A number of self-consistent calculations considering realistic boundary conditions are also available in the literature, which deviate considerably in describing the density profiles from the above mentioned analytical approaches.~\cite{Oh97:13519,Sefa08:prb,Ozge:contacts} Nevertheless, the basic idea is to minimize the total energy by the virtue of interactions to achieve an electrostatically stable density distribution, preserving boundary conditions. We will describe our density profiles utilizing results of the self-consistent calculations.
\begin{figure}[t!]
{\centering
\includegraphics[width=1\linewidth,angle=0]{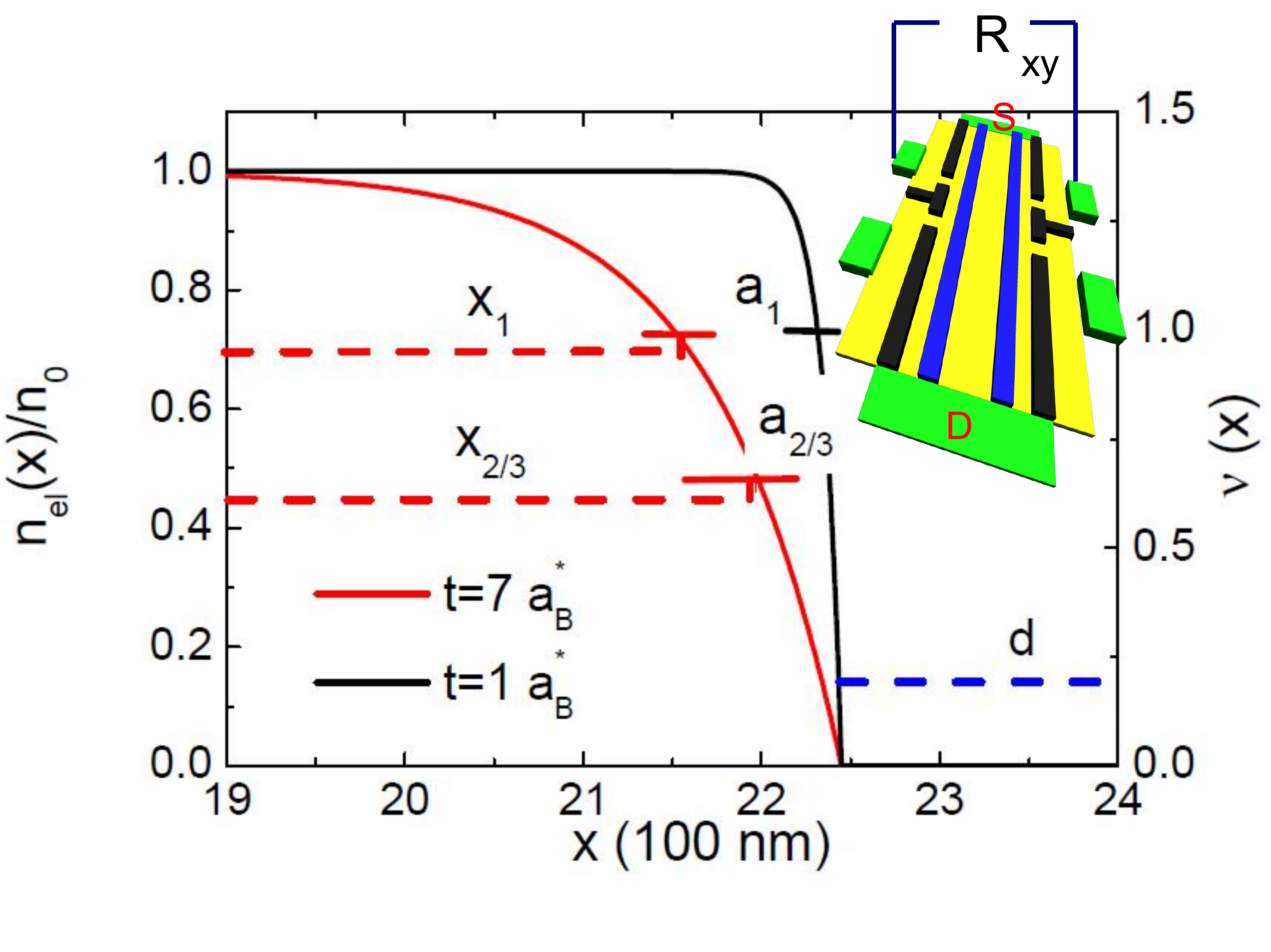}
\caption{The electron density distribution versus lateral coordinate, considering etched ($t=1~a_B^*$, steep profile, black) and gated ($t=7~a_B^*$, smooth profile, red) samples . Horizontal short lines denote the widths of incompressible strips ($a_i$) assuming $\nu=1$ and $\nu=2/3$, together with the central positions $x_i$. Inset: The sample geometry, black (broken) strips define the edge, whereas blue solid strips are the additional gates to smoothen the edge density. The width of the sample is $L$= 240 $a^*_B$ and depletion region is $d$= 20 $a^*_B$. Contacts are denoted by green regions, used as source (S), drain (D) and to measure transverse resistance $R_{xy}$.\label{fig:fig1}}}
\end{figure}
\subsection{Electron density model, finite wavefunctions and strip widths}
Here, we consider the geometry shown in the inset of Fig.~\ref{fig:fig1}, where the Hall bar is defined by side gates similar to Chklovskii. Light (yellow) region shows the plane of the 2DES, whereas the dark regions represent the gated areas. The contacts are denoted by light (green) regions. Applying a negative potential to the gates influences the electron density distribution, and leads to an inhomogeneous density profile. Since, the gating and etching processes affect the electron density distribution differently,~\cite{Chklovskii92:4026,Halperin94:etchedge,Oh97:13519,Sefa08:prb} we describe the distribution at $x$ by $n_{\rm el}(x)=n_0(1-e^{-(|x-|L-d||)/t})$ in the Hall bar region ($x<|L-d|$) and zero at the gated regions, motivated by the self-consistent calculations. The width of the sample is $L$. Here we take the bulk electron density $n_0\approx1\times10^{15}$ m$^{-2}$ and the depletion length $d$ of the order of few tens of effective Bohr radius $a_B^*$ ($=9.81$ nm for GaAs). Note that $t$ is a parameter which specifies the slope of the electron density profile, allowing us to describe both etched and gated samples. We set $t\gtrsim 3$ $a^*_B$ for gated samples and $t=1$ $a^*_B$ for etched samples, referring the numerical~\cite{Oh97:13519,Sefa08:prb} and analytical calculations.~\cite{Chklovskii92:4026,Halperin94:etchedge} Typical density distributions are shown in Fig.~\ref{fig:fig1}. The local filling factors at integral (or fractional) incompressible strips are described by $\nu(x_{k,f})=2\pi \ell^2_B n(x_{k,f})=\{k,f\}$, here $k$ $(=1,2,3..)$ stands for the integer states and $f$ $(=1/3,2/3..)$ for the odd denominator fractional states. The central position of the incompressible strips can be obtained from $x_{k,f}=|L-d|+t\ln(1-\{k,f\}/\nu_0)$, if $\{k,f\}<\nu_0$ condition holds. Here, $\nu_0=2\pi \ell_B^2n_0$ is the bulk filling factor.

\begin{figure}[t]
{\centering
\includegraphics[width=.9\linewidth,angle=0]{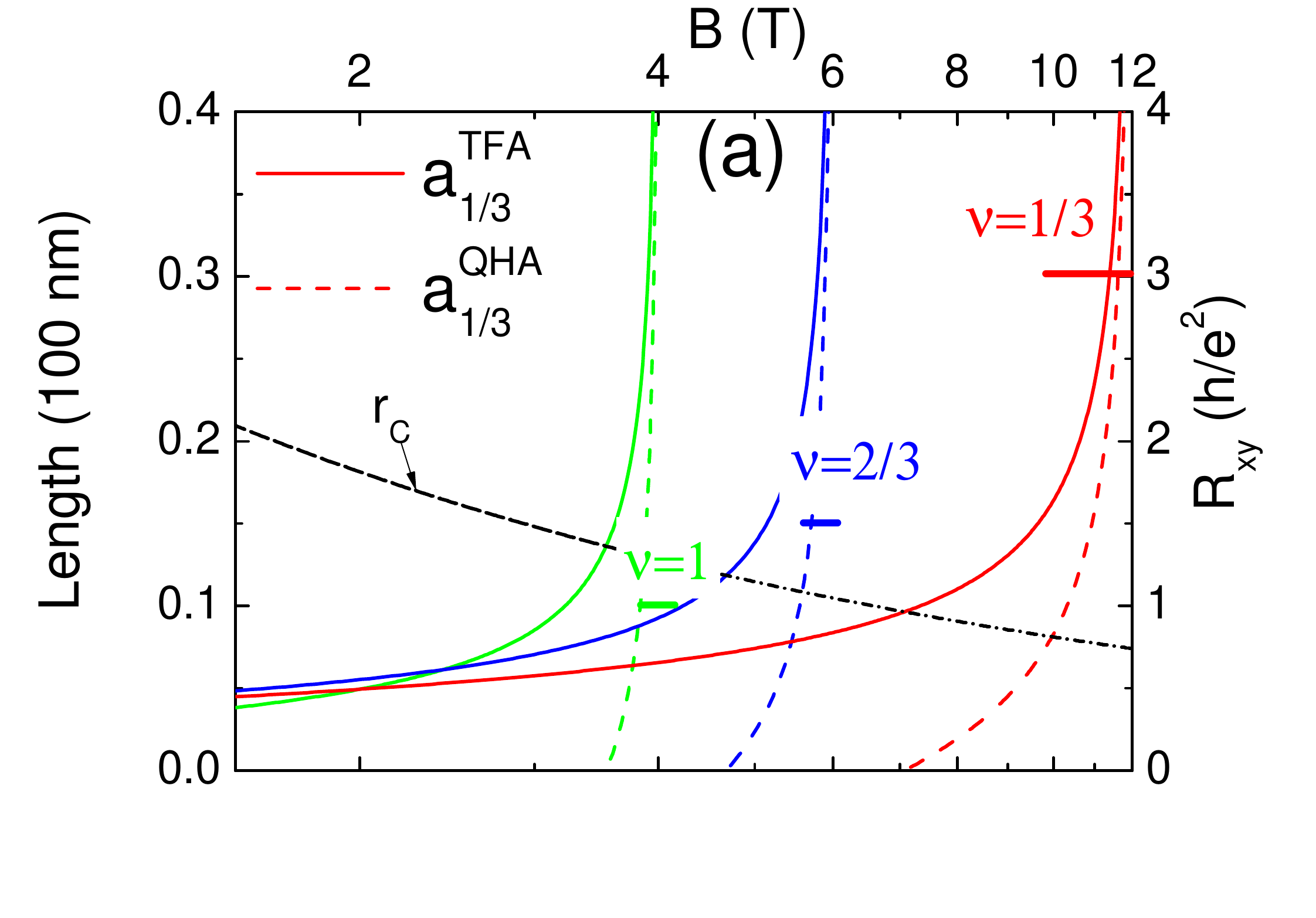}\vspace{-1.5cm}
\includegraphics[width=.9\linewidth,angle=0]{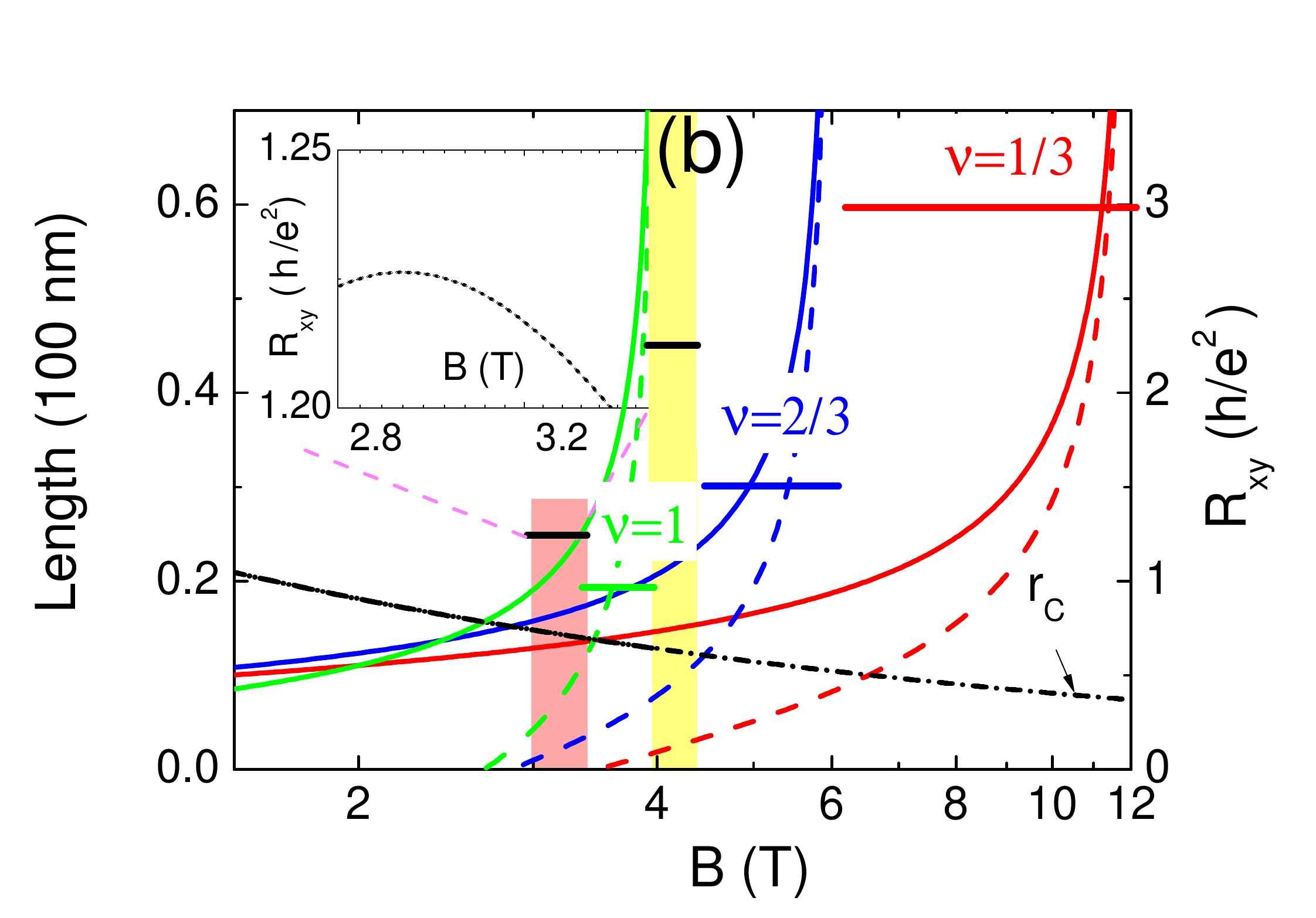}
\caption{The widths of ISs at some fractional fillings ($\nu=1/3, 2/3$) and at integer filling $\nu=1$ as a function of $B$ (left axis), together with the transverse resistance (solid horizontal lines, color coded) on the right axis. (a) Etched sample with $t=1 a_B^*$ and (b) gated sample with $t=5 a_B^*$, the black horizontal lines depict the maximal value of $R_{xy}$. Inset in (b) shows the calculated diagonal resistance at the overlap interval of $\nu=1$.\label{fig:fig2}}}
\end{figure}

The electrostatic stability condition of the current channels (incompressible strips) yields an analytical expression to calculate the widths of the strips with integer index,~\cite{Chklovskii92:4026}
\be a_k=\sqrt{\frac{2\epsilon\triangle{E}}{\pi^2 e^2 dn(x)/dx|_{x=x_k}}}\label{widths},\ee where, $\epsilon$ (=12.4 for GaAs/AlGaAs) is the dielectric constant and the derivative of the density $dn(x)/dx|_{x=x_k}$ should be calculated at the center of the strip. It is clear that the strip widths are proportional to energy gap $\triangle{E}$ between two adjacent quantized levels. In the original work, spin degree of freedom is neglected and the energy gap for all the integer states are assumed to be $\hbar\omega_c$, with the cyclotron frequency $\omega_c=eB/m^*$. Instead, we consider the Zeeman gap $g^*\mu_BB$ for the odd integer filling factors, enhanced by the exchange interactions.~\cite{Igor06:155314,afifPHYSEspin,GonulTFD:09} Taking into account Zeeman splitting and employing the non-self-consistent Thomas Fermi Approximation (TFA), together with the modified density profile, the widths of the integer strips read~\cite{siddiki10:10065012}
\begin{equation} \label{akwidths}
a^{\rm TFA}_k=\sqrt{\frac{4\alpha a^*_B}{\pi} \frac{t}{(\nu_0-k)}},
\end{equation}
where $\alpha(=\frac{g^* \mu_B B}{\hbar \omega_c})$ is a dimensionless strength parameter and gives the ratio of Zeeman energy to cyclotron energy. Due to the exchange interactions we set $\alpha=0.2$ to calculate the width of $k=1$ strip similar to other works,~\cite{Evans:93} which is justified by experiments.~\cite{Khrapai:05,Deviatov09:spe}Note that, TFA neglects the finite widths of the wavefunctions and is viable only for slowly varying potentials on the magnetic length scale. In contrast, the potential does not vary slowly under the condition $a_k^{\rm TFA}\lesssim \ell_B$ and TFA becomes invalid.~\cite{Suzuki93:2986} This situation is also carefully noted at the original work, however, is forgotten and left unresolved. The quasi Hartree approximation (QHA) is proposed as a solution for this situation,~\cite{siddiki10:10065012,siddiki2004} which includes the finite widths of wave functions by replacing delta functions of TFA with Landau wavefunctions. Meanwhile, the energy eigenvalues are described as in TFA. It is straightforward to show that the widths of incompressible strips, within the QHA, can be approximated as $a^{\rm QHA}_k=(1-\frac{\ell_{B}}{a^{\rm TFA}_k})a^{\rm TFA}_k$.~\cite{siddiki2004,siddiki10:10065012} The width of $k=1$ strip is shown in Fig.~\ref{fig:fig2}a, calculated within the TFA (solid lines) and QHA (broken lines) considering the etching (a) and the gate (b) defined samples. On one hand, once the $a^{\rm TFA}_k<\ell_B$ condition is reached, the cyclotron motion of the electron is no longer quantized, hence the classical Hall effect (CHE) is observed. In our plots we show the normalized cyclotron radius $r_C~(=R_C/\sqrt{2})$ instead of $\ell_{B}$, since we would like to use same length scale for integral and fractional states. On the other hand, if the $\nu=1$ strip becomes wider than the wave extent, the bulk and the opposing edges of the sample are decoupled and IQHE sets in, i.e. $a^{\rm QHA}_k>\ell_B$, unless $\nu_0<1$. The corresponding resistance is shown by horizontal thick line in the same figure, where the scale is given on the right vertical axis. The interesting case is the evanescent incompressible strip, for which the condition $a^{\rm QHA}_k<\ell_B<a^{\rm TFA}_k$ holds. In this situation, electrons can tunnel across the strip, back-scattering becomes slightly promoted and the Hall resistance deviates from its quantized value, accordingly. In what follows, we investigate the edge steepness conditions, in which integral and fractional evanescent incompressible strips co-exist. We show that, if co-existence takes place the transverse resistance will present anomalies (i.e. overshoot), similar to the integral case.~\cite{Overshoot:theo,Sailer:10}

\subsection{Fractional states}
In this subsection, we investigate the fractional states utilizing the well known composite fermion (CF) picture proposed by Jain.~\cite{Jain89} Similar to the treatment of Chklovskii et al., we calculate the widths of the fractional strips, within the electrostatic approach. In addition, we will also reconsider the finite extent of the composite fermion wavefunctions and use the density profile discussed above. A rigorous analytical treatment of the exchange and the correlation effects considering the formation of fractional strips is provided in Ref.~\onlinecite{Vignale:03(emp)}, where again a smoothly varying potential is considered within the TFA. There, the confinement potential is assumed to have a parabolic form and corresponding density profile is obtained. Due to the soft confinement, their calculations result in co-existence of many fractional states, which do not become evanescent at all. As discussed above, once the edge becomes steep and the finite widths of the wavefunctions are taken into account, the fractional strips are expected to collapse.~\cite{Chklovskii92:4026} Next we investigate the collapse and co-existence of evanescent incompressible strips in the fractional regime.

The relation between electron and CF filling factors is given by $\nu=\frac{\nu^{*}}{2p\nu^{*}\mp1}$, where $\nu^{*}$ stands for the filling factor of the CF and $p$ is an integer determining the order of fractional state, e.g. $p=1$ yields to state 1/3 for $\nu^{*}=1$. In this paper, we limit our discussion with the main fractional states, i.e. the $\nu(x_f) = f = 1/5, 1/3, 2/5, 2/3$ fractional filling factors. To calculate the strip widths for these fractional fillings we follow the literature to determine the energy gaps formed by the virtue of many body interactions.~\cite{fano86:2670,girvin85:581,moller05:045344} The energy gap expression for the fractional fillings, in a general form, can be given as $\Delta_{f}=c_f \frac{e^2}{\epsilon \ell_B}$, where $c_f$ is the coefficient to be determined by the corresponding filling factor. The value of $c_f$ for $f$=1/5, 1/3, 2/5 are calculated within different frameworks, which are given as $c_{1/5}=0.025,  c_{1/3}=0.104$ and $c_{2/5}=0.058$.~\cite{fano86:2670,girvin85:581,moller05:045344} The value of $c_{2/3}=0.104$ is taken from Ref.~[\onlinecite{moller05:045344}], where it is mentioned that ``particle-hole conjugate pairs should give precisely the same excitation energies". There $\nu=1/3$ and $\nu=2/3$ states are considered as particle-hole pair of each other. We would like to mention also the earlier works by Haldane~\cite{Haldane:85} and Morf~\cite{Morf:86}, where the finite size effects are investigated in the former and a disc geometry is considered in the latter. All the cited works, predict approximately same gaps and the sufficiently small differences do not influence our results on the plot scales we use. Substituting the energy gap to the strip width expression given above, one obtains
\begin{equation} \label{afwidths}
a^{\rm TFA}_f=\sqrt{\frac{4 \ell_B c_f}{\pi} \frac{t}{(\nu_0-f)}}
\end{equation}
for the fractional strips. Note that, this expression differs from the definition of Chklovskii et al. by a pre-factor of $\sqrt{2}$ which enables us to compare the normalized cyclotron radius of both electrons and CF on the same plot.

Similar to the integer case we calculate the widths of the fractional strips using the above introduced density profile and approximations. In Fig.~\ref{fig:fig2}, the width of the strips calculated within TFA ($a_{k,f}^{\rm TFA}$) and QHA ($a_{k,f}^{\rm QHA}$) are shown by solid and broken lines respectively. As mentioned before, once the wave width of the composite fermion $\ell_{CF}~(=r_C)$ becomes larger than the strip width, i.e. $a^{\rm QHA}_f>\ell_{CF}$, incompressible strip decouples edges and the fractional Hall plateau is observed. Turned around, the spatial extent of the many-body effect induced energy gapped region is large to decouple the probe contacts. In contrast, $a^{\rm TFA}_f<\ell_{CF}$, assures the CHE. The strip is so narrow that it cannot wrap up the composite fermion cyclotron radius, hence many-body gap collapses. In the interval $a^{\rm TFA}_f>\ell_{CF}$$\gtrsim$ $a^{QHA}_f$ the current flows through the strip and \emph{a} coherent electronic state resides inside this \emph{evanescent} incompressible strip (eIS).~\cite{Sailer:10,siddiki10:10065012} In this situation, electrons can tunnel across the strip, hence it is no longer a well defined incompressible state. Therefore, the excess current can be redistributed between evanescent incompressible strips with different filling factors, if they co-exist. Next, we seek for magnetic field intervals where these eIS co-exist similar to the integer case.~\cite{Sailer:10,siddiki10:10065012,Overshoot:theo}

It is worth to note that, our above arguments should be modified if one considers i) temperature effects and ii) the effects arising from local electric fields. Most importantly, iii) the length scale $\ell_{CF}$ depends on the answer to the question ``how many electrons can be considered as a composite fermion ?'', or explicitly, how many electrons should reside within the incompressible strip to have a well developed fractional state. Our understanding is as follows: i) we consider absolute zero for temperature, which is plausible if the experiments are performed at dilution fridges below 0.5 K ii) The effects that we discuss in what follows can be observed under (approximately) equilibrium conditions. Even then, the narrow strips (e.g $a^{\rm QHA}\approx \ell_{CF}$) can smeared by local electric fields.~\cite{SinemLDOS10:,TobiasK06:h} iii) Here, we rely on the basic arguments of the literature.~\cite{Chklovskii92:4026,Jain07:priv,AdyStern:quantumcomp} In addition, the translational invariance lifts the restriction on the area that the strip can accommodate sufficient number of electrons. Moreover, as we will show below, even if the length scale is taken to be larger up to a factor of 1.5-2, the effects we consider are still observable for sufficiently smooth edge density profiles (e.g. $t\geq7~a_B^*$).
\section{Co-existence and interference}
We start our discussion considering an etched defined sample, $t=1~a_B^*$. The typical strip widths are shown in Fig.~\ref{fig:fig2}a, focusing on the integer ($\nu=1$, the left-most pair of curves) and principle fractional state(s), $\nu=1/3$ and 2/3. In this sample design, the plateaus are well developed (depicted by horizontal short-lines, color coded) once the strip widths calculated within the QHA is larger than the magnetic length of the corresponding particle. Since, the edge potential is considerably steep, the incompressible strips fade away as soon as the plateau ends. Evanescent incompressible strips with different filling factors do not co-exist. In Fig.~\ref{fig:fig2}b, we show the co-existence intervals by the shaded areas. We observe that, the $\nu=1$ and $\nu=2/3$ evanescent incompressible strips co-exist in the interval $3\lesssim B\lesssim 3.5$ T (the darker shaded area), for the gate defined sample. However, for the etched sample no co-existence is observed (Fig.~\ref{fig:fig2}a) since the derivative in Eq.~\ref{widths} is large. The eIS co-existence is also observed when considering $\nu=1/3$ and 2/3 fractional states for the gated sample.

So far we investigated the conditions to observe co-existence of evanescent incompressible strips with distinct filling factors, and depicted them in Fig.~\ref{fig:fig2}. However, in usual transport experiments only the global resistances are measured. The widths of the strips can only be measured by local probe experiments.~\cite{Ahlswede02:165,Franck:10} Therefore, we should estimate the total transverse resistance $R_{xy}$, due to co-existence. If we assume that the excess current $I$ is fixed and is shared among the eISs proportional to their widths, then the total resistance across the bar is given by,
\be R_{xy}= \sum_{i=k,f} \frac{a_i}{a_T}.R_i,\label{resistance}\ee
where $i$ runs over integer and fractional states and $a_T$ is the total width of the evanescent incompressible strips. A semiclassical Boltzmann approach can be used to describe transport at these channels,~\cite{Zwerschke99:2616} however, in this work we are not concerned with a detailed transport formulation. We use the commonly used fractional quantized conductance (namely, resistance). If the current is equally distributed to the strips, total resistance is nothing but the sum of individual channels. Let us consider the case in Fig.~\ref{fig:fig2}b regarding $\nu=1$, $\nu=2/3$ channels and assuming $a_1=a_{2/3}$, then $R_{xy}=\frac{h}{e^2}[(1/2).1+(1/2).(3/2)]=(5/4)\frac{h}{e^2}$. The corresponding resistance is denoted by thick horizontal line in the co-existence region. However, the total resistance within the co-existence (or overlap) interval varies due to the fact that the widths of the channels also vary depending on the field strength. A back of envelope calculation results in the curvature shown in the inset of Fig~\ref{fig:fig2}b. Once the outer-most strip fades away, the transverse resistance ends in the classical value, i.e. $R_{xy}\sim\nu\frac{h}{e^2}$. This is the overshoot of fractional states and is affected by i) temperature, ii) the curvature of the density and iii) by the mobility of the sample (namely, long-range fluctuations caused by disorder). We ought to observe fractional overshoot only at temperatures of few hundreds of mK and at high mobility ($>2\times10^6$ Vm/s) gate defined samples.
\begin{figure}[t!]
{\centering
\includegraphics[width=.9\linewidth,angle=0]{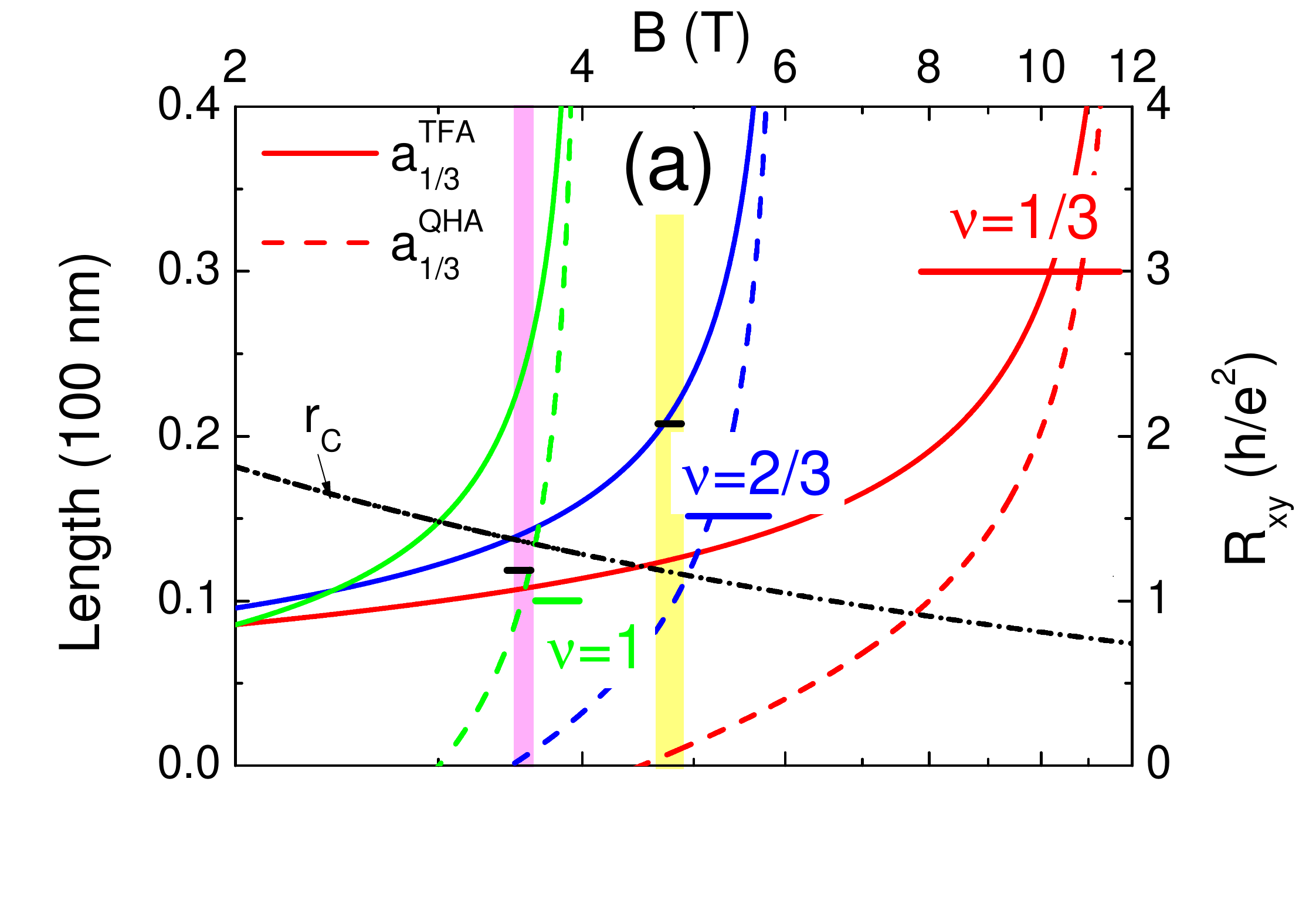}\vspace{-1.5cm}
\includegraphics[width=.9\linewidth,angle=0]{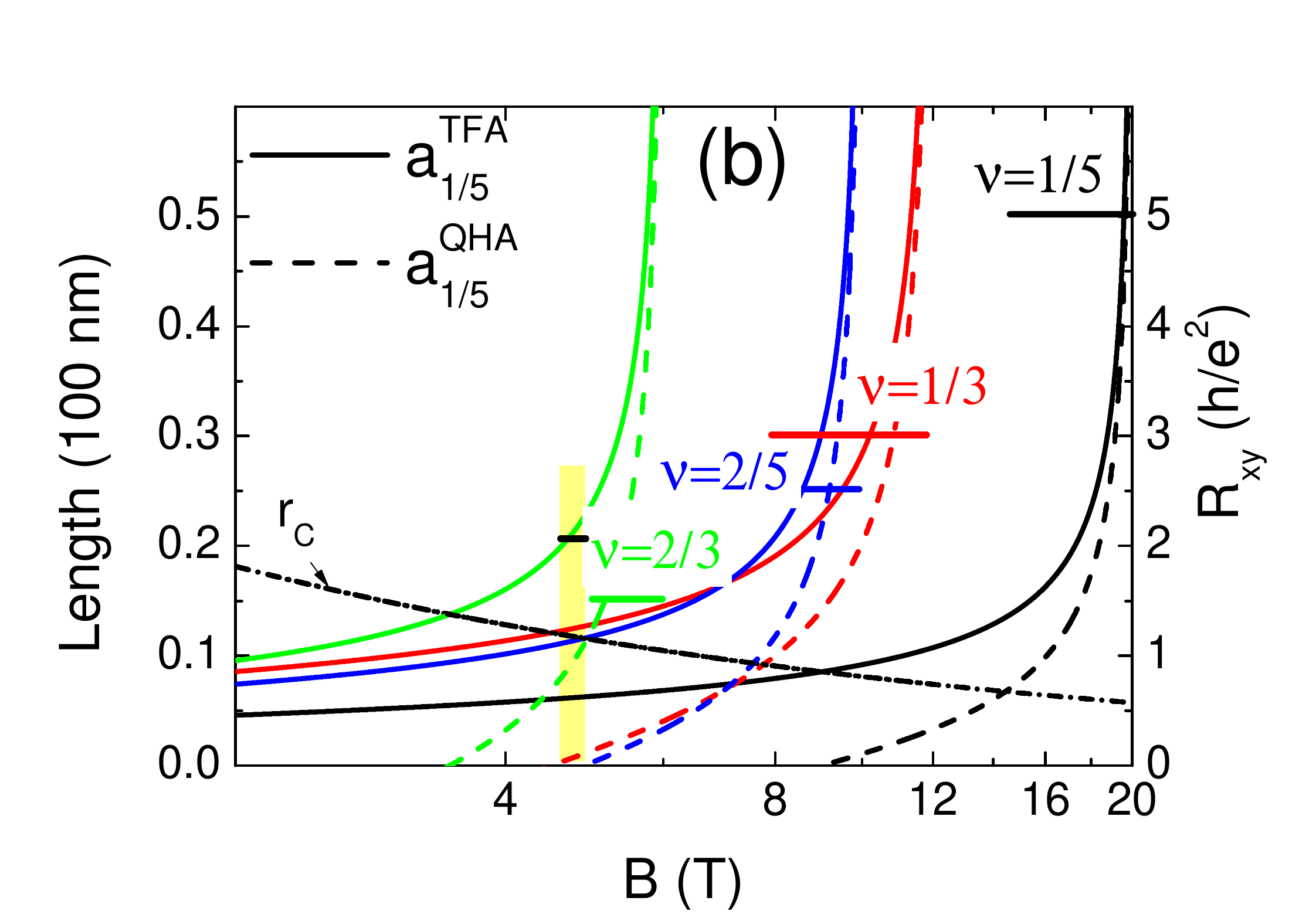}
\caption{The widths of incompressible strips calculated for $\nu=1, 1/5, 1/3, 2/5,$ and $2/3$ as a function of $B$. For $t=3~a_B^*$.\label{fig:fig3}}}
\end{figure}
\begin{figure}[t!]
{\centering
\includegraphics[width=.9\linewidth,angle=0]{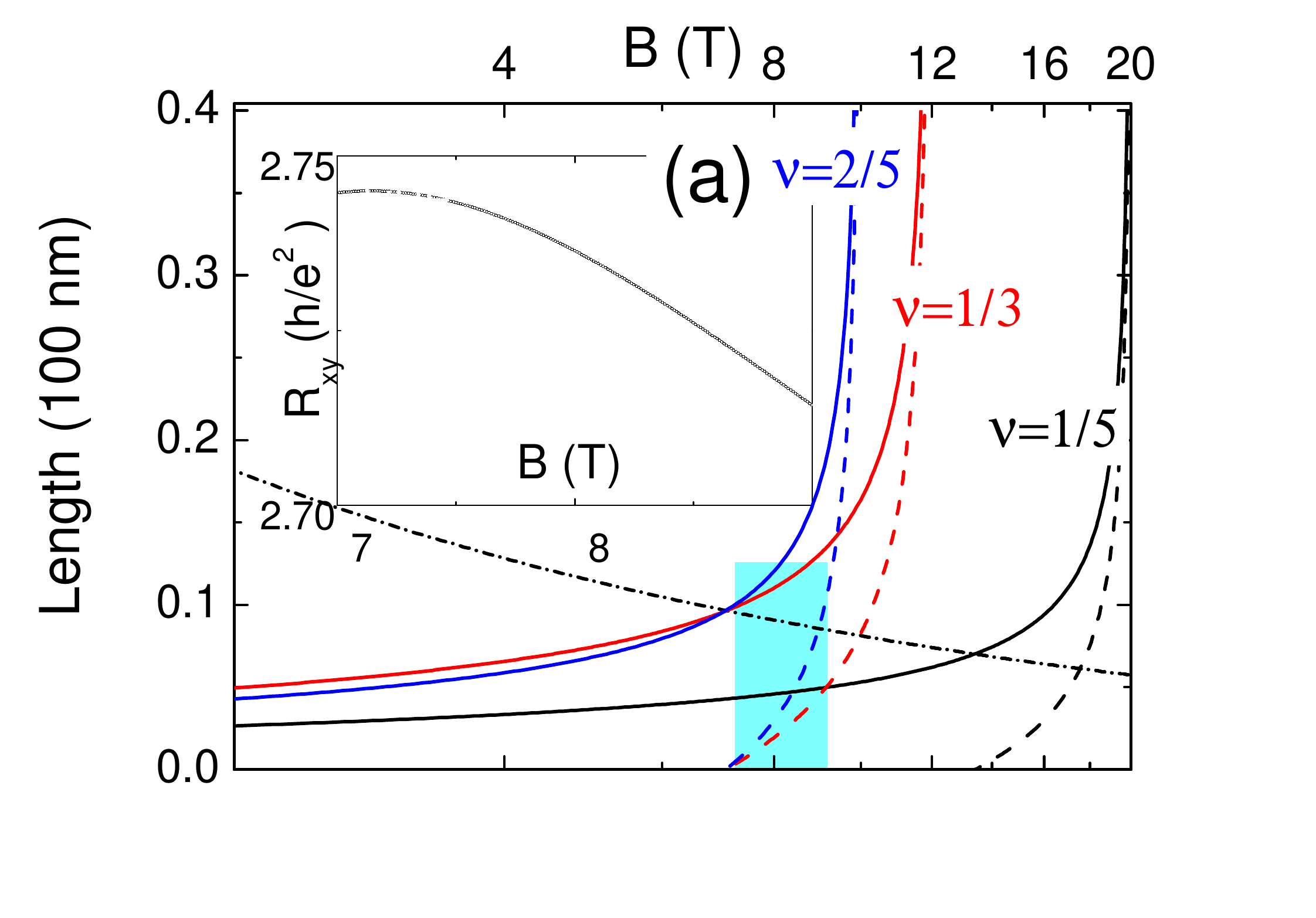}\vspace{-1.5cm}
\includegraphics[width=.9\linewidth,angle=0]{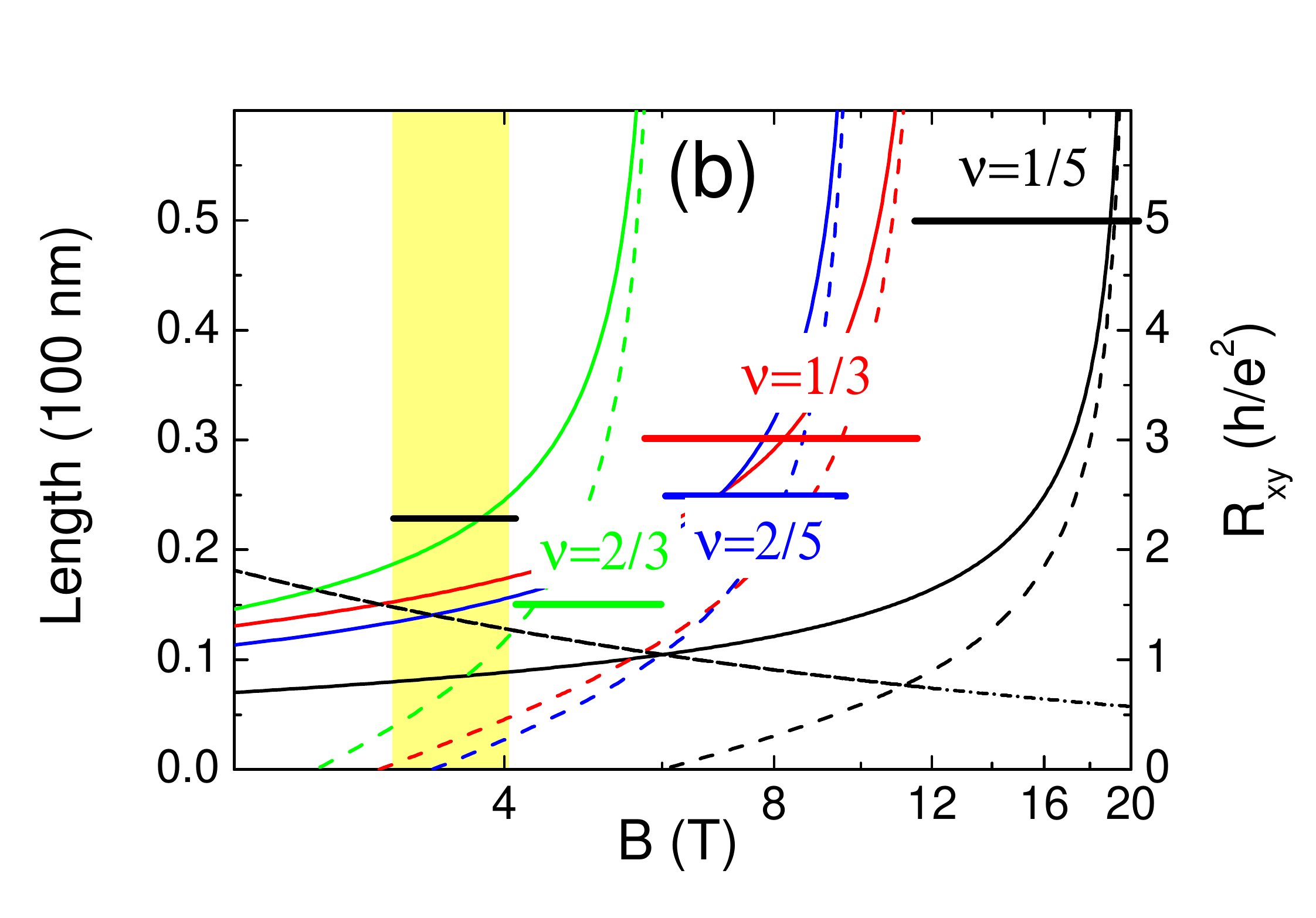}
\caption{The widths of incompressible strips calculated for $\nu=1/5, 1/3, 2/5,$ and $2/3$ as a function of $B$. (a) Considering an etched sample with $t=1~a_B^*$ and (b) gated sample with $t=7~a_B^*$.\label{fig:fig4}}}
\end{figure}

We observed that, overshoot intervals show prominent differences considering etching and gating processes, for the main fractions such as 1/3 and 2/3. We proceed our investigation with higher odd-integer denominator fractional states and consider an intermediate edge steepness, namely $t=3$ $a_B^*$. Notably, at this particular steepness one starts to observe the co-existence of integral and fractional states. Fig.~\ref{fig:fig3}a depicts the results of our calculations considering such a sample and regarding the primary fractions only. Here co-existences of $\nu=1-2/3$ and $\nu=2/3-1/3$ states can be observed. Taking into play also the secondary fractions, i.e. $\nu=1/5,2/5$, the picture is modified as shown in Fig.~\ref{fig:fig3}b. Strikingly the $\nu=1/3$ plateau covers the 2/5 state, even at the bulk, which makes us to conclude that for gate defined samples observation of 2/5 state becomes hindered. The experimental evidence for such a situation is given as follows: The Fabry-Perot interference setup is obtained by a modification of the Corbino geometry, that is a circular Hall configuration with a hole at the center, and side gates are operated to manipulate the transport at the edges. Electrons are forced to tunnel across the edge incompressible strip and the current is measured as a function of external magnetic field at different filling factors. Detailed description of the experiments can be found in Ref.~[\onlinecite{Deviatov09:fabry}]. An interference signal is measured for $\nu=1,2/3$ and 4/3 states, however, it is observed that the principle Laughlin state (i.e. $\nu=1/3$) is missing. This effect is attributed to interference of ``normal'' electrons, since 4/3 and 2/3 states are regarded as excitations of electron or hole states. Although, we agree that the interference is due to normal electrons, our results propose that the 1/3 state is missing due to the strong 2/5 bulk state. This bulk state hinders the interfering electrons to generate an oscillation at the conductance.

Considering the etched sample in Fig.~\ref{fig:fig4}a we observe that, $\nu=1/3$ and $\nu=2/5$ strips co-exist. For gated samples (b) $a^{\rm TFA}_{1/3}$ emerges before $a^{\rm QHA}_{2/5}\lesssim \ell_{CF}$ condition sets in. This observation also coincides with the experimental findings of Camino \emph{et al},~\cite{camino05:075342, camino06:115301} pointing that the $\nu=1/3$ is the edge state, whereas $\nu=2/5$ is the bulk state. In these experiments the Aharonov-Bohm phase of quasi-particles are measured precisely, and the conductance oscillations suggest that a 1/3 Laughlin quasi-particle encircles the 2/5 bulk state. Form the resistances point of view, if the bulk 2/5 state dominates then the total resistance approaches to 2.5 $h/e^2$, whereas if the edge 1/3 state is as wide as the 2/5 state the resistance reads $R_{xy}\approx2.75$ $h/e^2$. This edge-bulk separation also brings a constraint on observing 2/5 state, since if the 1/3 edge state decouples the bulk, one cannot observe $\nu=2/5$ plateau. Our results point that, one has to define physical edges of the system in an extremely sharp way. Such an edge profile can be either obtained by deep etching or by trench-gating, where the metallic gates are deposited after the etching process. Similar to the experiments.~\cite{Camino05:075342} The hindering of 2/5 state by 1/3 becomes more evident if one considers a gated sample as in Fig.~\ref{fig:fig4}b. Here, the $\nu=1/3$ plateau exceeds the full interval of $\nu=2/5$ plateau. In such a gated sample one also strikingly observes that, 2/3, 2/5 and 1/3 evanescent incompressible strips overlap all together. A plausible estimate of $R_{xy}$ is about 2.15 $h/e^2$, depicted in the interval by (yellow) shaded area.
\begin{figure}[t]
{\centering
\includegraphics[width=.9\linewidth,angle=0]{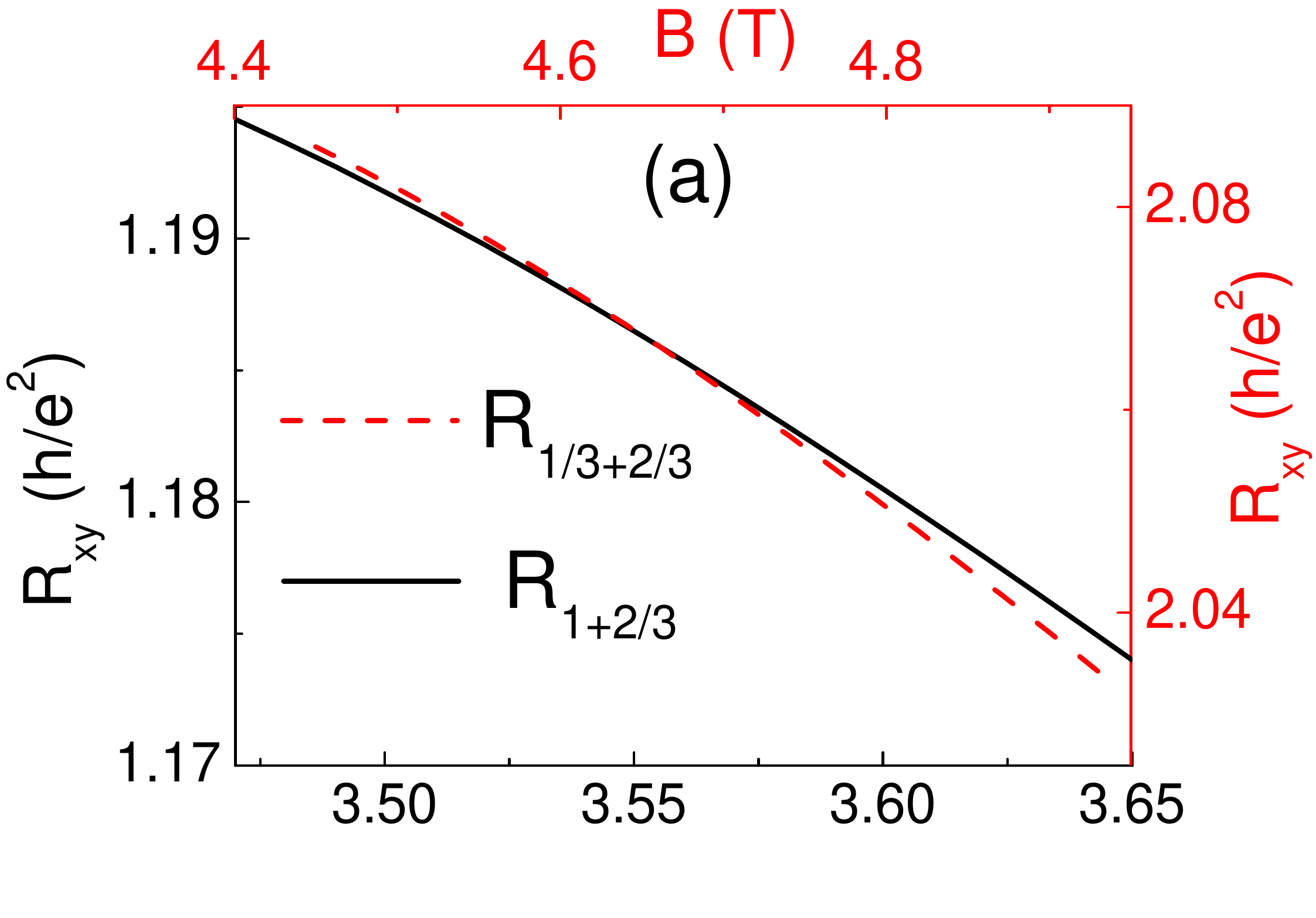}
\includegraphics[width=.9\linewidth,angle=0]{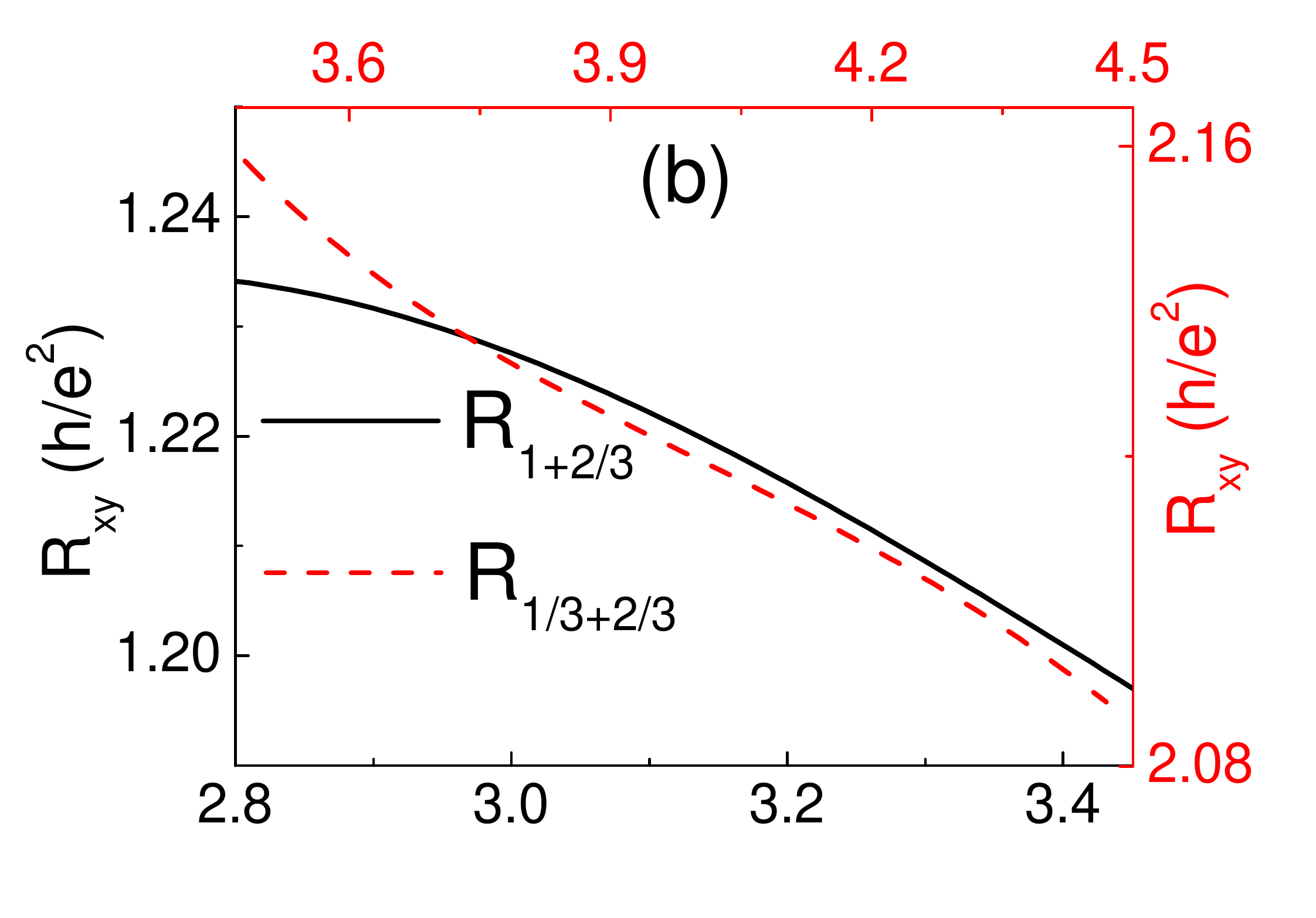}\vspace{-.0cm}
\includegraphics[width=.9\linewidth,angle=0]{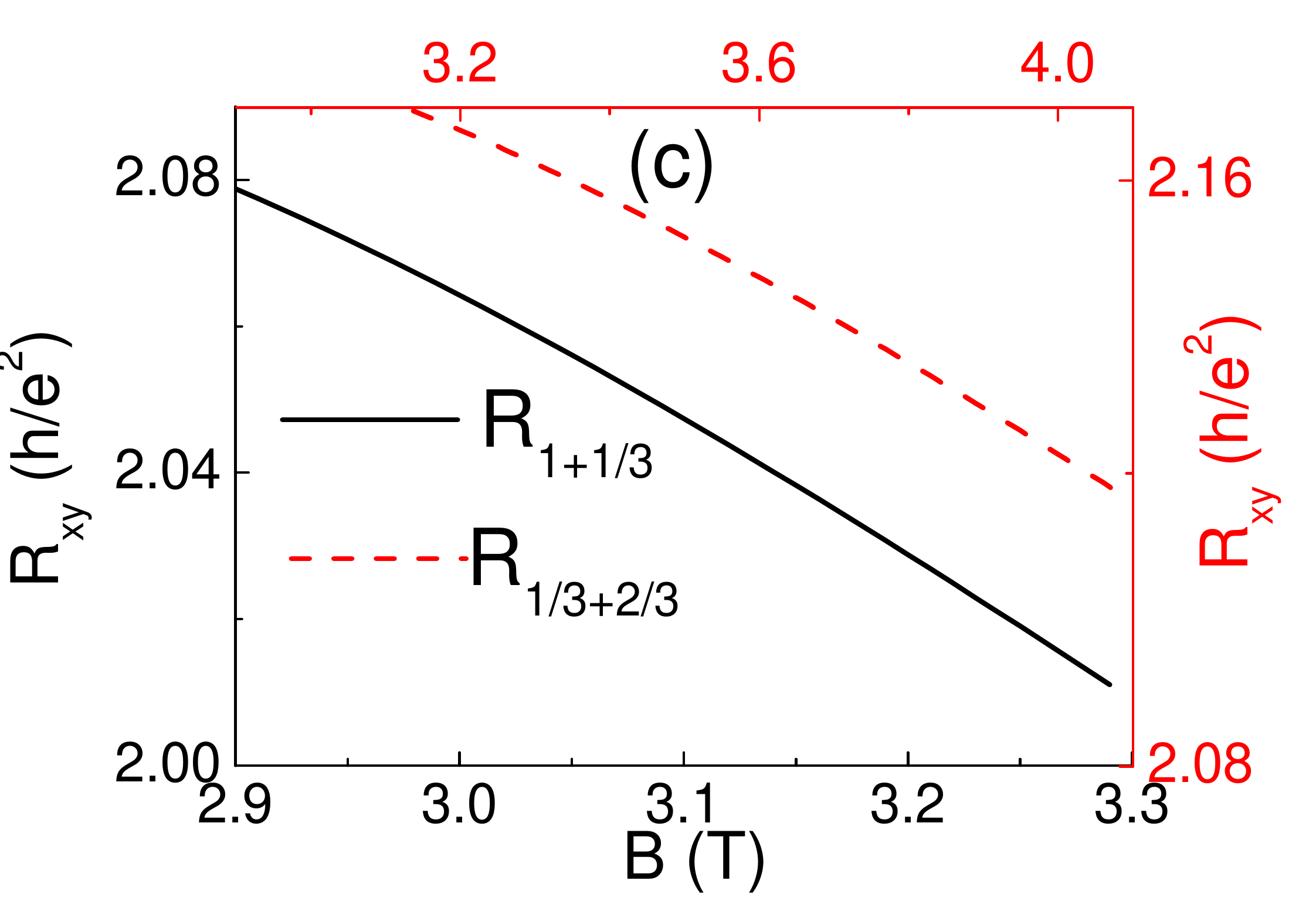}
\caption{The transverse resistances calculated for the overlaps $\nu=1/5, 1/3, 2/5,$ and $2/3$ as a function of $B$. (a) $t=3~a_B^*$ and (b) gated sample with $t=5~a_B^*$, (c) $t=7~a_B^*$.\label{fig:fig5}}}
\end{figure}
Fig.~\ref{fig:fig5}, depicts more careful calculations of the transverse resistances taking into account strip widths according to Eq.~\ref{resistance}. Depending on the strip widths and resistances of each channel, the Hall resistance overshoot presents linear (Fig.~\ref{fig:fig5}a) or sub-linear (Fig.~\ref{fig:fig5}b-c) behavior as a function of the field strength. Interestingly enough, the $R_{xy}^{1/3-2/3}$ exhibits an increase stepper than other co-existence combinations for $t=5~a_B^*$, Fig.~\ref{fig:fig5}b. These results show that, if one sweeps the steepness of the edge profile by changing the side gate voltages the overshoot resistance behaves strongly non-linear. To investigate the effect of edge steepness, we calculated the overshoot conditions and intervals as a function of $t$ as depicted in Fig.~\ref{fig:fig6}. The upper panel, presents the expected magnetic field intervals of evanescent incompressible strips of 1-1/3 (left), 2/3-1/3 (middle) and 2/5-1/3 (right) while varying the gate voltage, i.e. $t$. Horizontal dashed lines constrain the field strengths, if a larger length scale ($\sim 2*\ell_{B}$) is considered. It is observed that, the expected $B$ values of the evanescent incompressible strips decreases by increasing $t$ for all states. Due to the fact that, the density also decreases. Remarkably, the decreasing rates are not same for all states, resulting in a non trivial co-existence interval as shown in the lower panel of Fig.~\ref{fig:fig6}. It is apparent that, while increasing $t$ (namely by making the edge smoother), overlap interval increases for 1-2/3 and 2/3-1/3 co-existences, however, starts to decrease at $t\sim4~a_B^*$ for 2/5-1/3. This result essentially presents the hindering of bulk 2/5, by 1/3 edge incompressible strip, discussed above to elucidate Fabry-Perot experiments.

It is important to note that, the actual size of the CF particle is in principle unknown. Here, in our model we therefore investigated different length scales, namely the magnetic length $\ell_B$, the cyclotron radius of electron or composite fermion and the extent of the wavefunction $\sim\ell_B$ or $\ell_{CF}$, to compare with the width of the corresponding incompressible strip. The length scales have the same order of magnitude up to a pre-factor of $\sqrt{2}$, at least for the lowest Landau level. In addition, the widths of the strips are commonly larger than the theoretical estimates, due to the fact that unscreened Hartree-Fock approximation corrected by correlations usually consider extremely smooth (and in some unrealistic) confinement potentials. Even if, the strip widths become sufficiently narrow by the virtue of steep confinement, it is possible to vary the steepness by manipulating the gate potentials. All summed up, we expect to observe the above discussed co-existence of the evanescent fractional incompressible strips that yield an overshooting at the transverse resistance.
\begin{figure}[t]
{\centering
\includegraphics[width=.9\linewidth,angle=0]{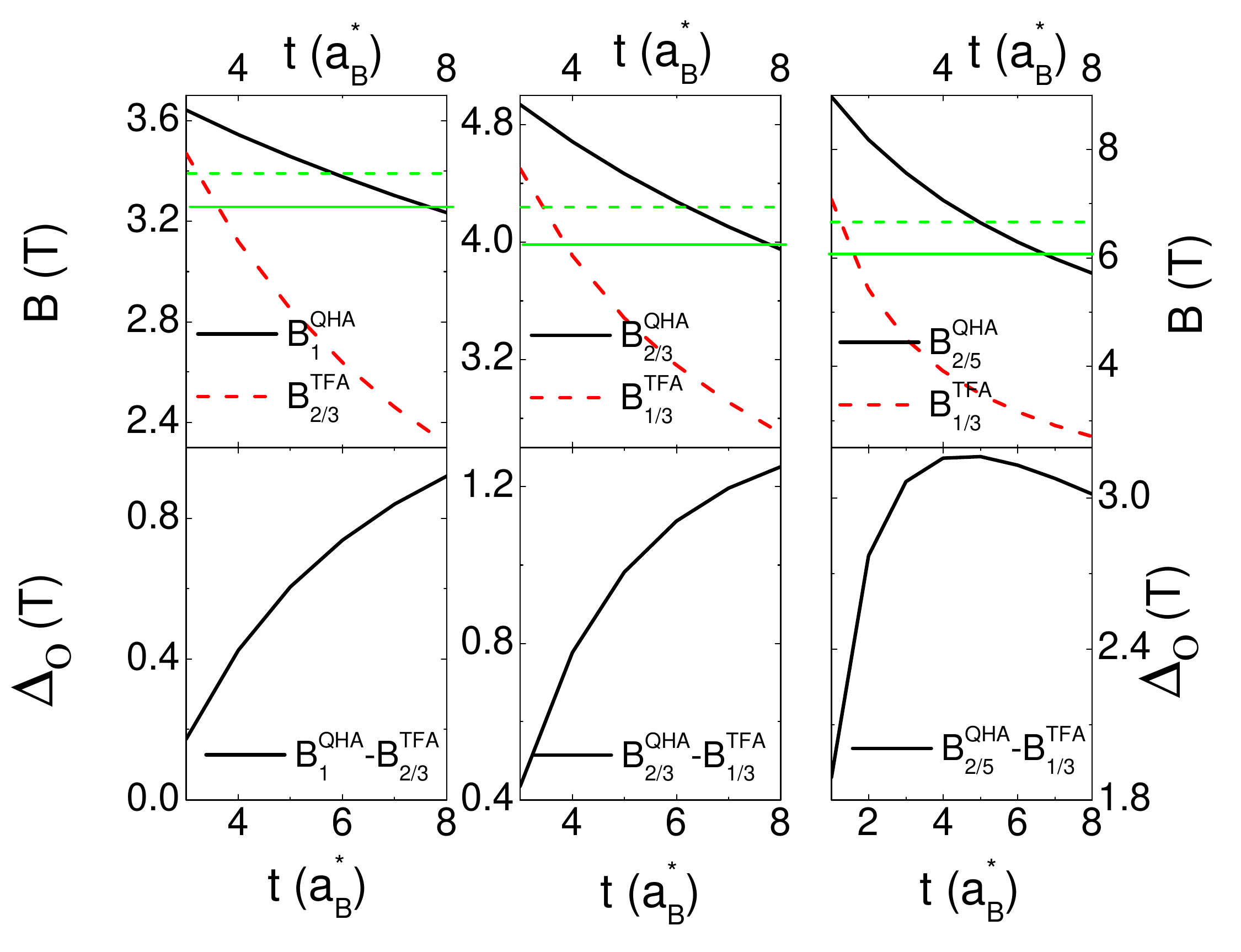}
\caption{The calculated overshoot intervals, considering different filling factor combinations.\label{fig:fig6}}}
\end{figure}

\section{CONCLUSION}
The above analytical calculations considering main fractional states via electrostatics, show that one should observe a similar behavior to the overshooting effect in the fractional domain, similar to the integral one. Although, the edge dynamics is intensively studied in experiments~\cite{oto:97,Ernst:97}, up to our knowledge, no systematic investigation of the transport measurements are performed regarding the gate defined narrow samples. Whereas, similar samples are used to investigate the incompressible strips at the edge in the integral quantized Hall effect regime.~\cite{afif:expas} We found that the smoothness of the edge profile generates complications due to co-existence of well developed fractional incompressible strips. In addition we show that, it is possible to end in a situation, where observation of the bulk fractional state is hindered. Since our calculations are limited to $T=0$, the effects discussed above can only be observed at  temperatures below  0.5 K and are very sensitive to external parameters, such as current amplitude etc. To protect the fractional overshooting, we suggest to deposit additional metallic gates on the surface parallel to the edges, as depicted in Fig.~\ref{fig:fig1}. Applying a negative potential to the additional gates (smaller than edge defining gates) will result in reduction of electron density beneath, hence the fast density variation at the edges can be suppressed. Such a smooth density variation will allow us to protect the smearing of outer-most eIS. Therefore, one can obtain more than one eIS by varying the top gate voltage. In principle the width of the gates should be sufficiently narrow (e.g. $<100$ nm) to avoid variations at the bulk density.

In summary, we employed the non-self-consistent electrostatic picture of Chklovskii et al in the fractional regime with corrections to the density profile and also considering finite width of the wavefunctions. We used the energy gaps for the fractional states, obtained from composite fermion picture. We have shown that, if the edge density profile is sufficiently smooth one can obtain the condition $a^{\rm TFA}_{k,f}>\ell_B>a_{k,f}^{\rm QHA}$ (i.e. the evanescent incompressible strip) for more than one state. In such a situation, the total transverse resistance increases depending on the widths of the overlapping strips, resembling the overshooting effect of integer quantized Hall effect. In particular, we found that the co-existence of $\nu=2/5$ and $\nu=1/3$ states may result in the hindering of the inner bulk state. Such a behavior elucidates the unexpected behavior of missing principle Laughlin state at the Fabry-Perot interference experiments. Finally, we proposed a yet un-investigated sample design to enhance the overshooting effect of fractional states utilizing additional surface gates.

\begin{acknowledgments}
We would like to thank M. Grayson and A. Wild for discussions on the overshooting effect and J. Jain for his initiating ideas on utilizing the composite fermion picture in our calculations and A. Erzan for her critical questions on this picture, which helped us in improving our understanding. This work is financially supported by the scientific and technological research council of Turkey (T\"UB\.ITAK), under grant no: 109T083 and Istanbul University BAP:6970.
\end{acknowledgments}


\end{document}